\documentclass[11pt,sigplan,screen]{acmart}

\AtBeginDocument{%
  \providecommand\BibTeX{{%
    \normalfont B\kern-0.5em{\scshape i\kern-0.25em b}\kern-0.8em\TeX}}}

\setcopyright{acmlicensed}
\copyrightyear{2024}
\acmYear{2024}
\acmDOI{XXXXXXX.XXXXXXX}

\acmConference[HotEthics'24]{Make sure to enter the correct
  conference title from your rights confirmation emai}{April 28, 2024}{San Diego, California, USA}
\acmISBN{978-1-4503-XXXX-X/18/06}

\begin{document}

\title{Towards Privacy-Preserving Audio Classification Systems}

\author{Bhawana Chhaglani}

\affiliation{%
  \institution{University of Massachusetts Amherst}
  \city{Amherst}
  \country{USA}
}
\email{bchhaglani@umass.edu}

\author{Jeremy Gummeson}
\affiliation{%
  \institution{University of Massachusetts Amherst}
  \city{Amherst}
  \country{USA}
}
\email{jgummeso@umass.edu}

\author{Prashant Shenoy}
\affiliation{%
  \institution{University of Massachusetts Amherst}
  \city{Amherst}
  \country{USA}
}
\email{shenoy@cs.umass.edu}

\renewcommand{\shortauthors}{Chhaglani et al.}

\begin{abstract}
  Audio signals can reveal intimate details about a person's life, including their conversations, health status, emotions, location, and personal preferences. Unauthorized access or misuse of this information can have profound personal and social implications.  In an era increasingly populated by devices capable of audio recording, safeguarding user privacy is a critical obligation. This work studies the ethical and privacy concerns in current audio classification systems. We discuss the challenges and research directions in designing privacy-preserving audio sensing systems. We propose privacy-preserving audio features that can be used to classify wide range of audio classes, while being privacy preserving.

  

\end{abstract}

\ccsdesc[500]{Human-centered computing~Personal digital assistants}
\ccsdesc[500]{Security and privacy}

\keywords{Audio classification, Privacy-preserving sensing, Acoustic features}


\received{20 February 2007}
\received[revised]{12 March 2009}
\received[accepted]{5 June 2009}

\maketitle

\section{Introduction}

Audio sensing has tremendous potential and can be used to infer a rich array of personal information about an individual \cite{zhang2019pdvocal,swain2018databases,chhaglani2022flowsense,chhaglani2023cocoon,chhaglani2024aerosense,chhaglani2023breatheasy}. Kroger et al. \cite{kroger2020privacy} presents an overview of sensitive pieces of information that can, with the help of advanced data analysis methods, be derived from audio, including cues to a speaker’s biometric identity, personality, physical traits, geographical origin, emotions, level of intoxication and sleepiness, age, gender, and health condition. These are all examples of privacy-sensitive information and its access should be regulated.
Today, we are surrounded by listening devices including smart speakers, voice assistants, baby monitors, smoke alarms, and other emerging device categories. Some of them are always-on and listening to sensitive content, while others can be accidentally triggered \cite{schonherr2022exploring}. Thus, privacy remains one of the core problems in audio sensing which needs to be addressed.  

\begin{figure}
    \centering
    \includegraphics[width=0.8\columnwidth]{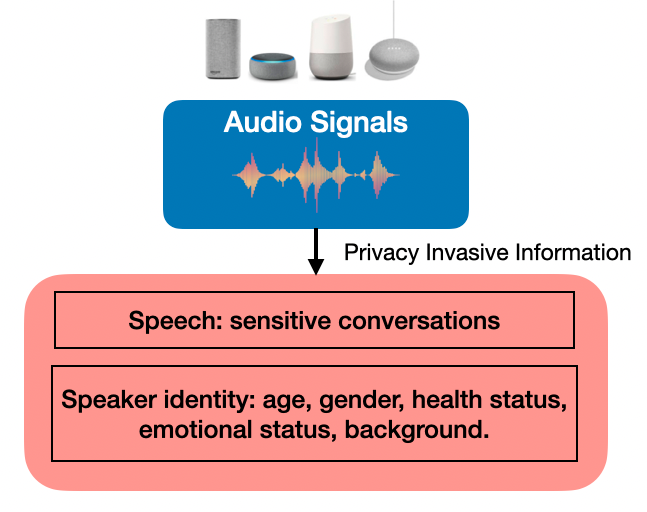}
    \caption{Privacy Concerns in Audio Classification Systems}
    \label{fig:privacy}
\end{figure}


Existing audio classification systems use features like spectrograms that can potentially compromise user privacy. To avoid privacy concerns, prior work has primarily focused on eliminating or obfuscating speech to ensure privacy \cite{xia2020pams,chen2008audio}. However, speech is not the only privacy sensitive information present in the audio signal --- it can also reveal information about the speaker's identity and whereabouts. The idea is to explore a more holistic definition of privacy and propose a mechanism to evaluate and quantify the privacy of an audio-based system. Through this research, we aim to explore the potential of audio sensing in enabling novel applications, while simultaneoulsy ensuring that user privacy is retained. 

In this work, we discuss the key challenges in realizing privacy-aware audio classification systems. Using the example of environmental sound classification, we demonstrate the application of privacy-preserving audio features to safeguard privacy. We show that this approach is generalizable and can achieve comparable accuracy as the state-of-the-art techniques that do not consider privacy. 


\section{Need for Ensuring Privacy in Audio Classification}
\begin{figure}
    \centering
    \includegraphics[width=0.99\columnwidth]{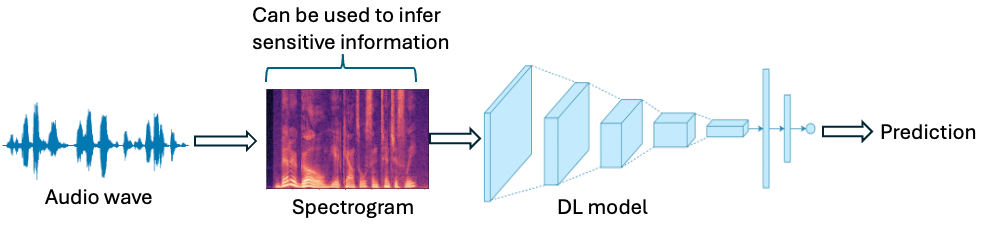}
    \caption{Typical Audio Classification Pipeline}
    \label{fig:pipeline}
\end{figure}

Most audio classification systems convert audio signals into Mel-Frequency Cepstral Coefficients (MFCCs), Mel spectrograms, and Short-Time Fourier Transform (STFT) and then feed them into advanced deep learning models as shown in Figure \ref{fig:pipeline}. 
Piczak et al. \cite{piczak2015environmental} explores the use of Mel spectrograms as input to convolutional neural networks (CNNs) for the classification of environmental sounds. Forsad et al. \cite{al2020flusense} compares MFCC and STFT for cough with CNNs for detection task. By effectively bridging the gap between raw audio signals and the sophisticated pattern recognition capabilities of deep learning (DL) architectures, these features enable the models to achieve remarkable accuracy and efficiency in variety of tasks. 
However, the same features that enhance the performance of audio classification models can also pose significant privacy risks. MFCCs, STFT, Mel spectrograms or other spectrograms, by their very nature, encapsulate detailed information about the audio source, potentially including sensitive personal information. For instance, MFCCs can retain distinctive speech characteristics that could be exploited to identify a speaker, thereby leaking personal identity information. Similarly, STFT and Mel spectrograms can inadvertently reveal background noises or conversations that were not meant to be shared, thereby compromising the privacy of individuals. The practice of transmitting spectrograms to cloud-based services for model prediction further heightens privacy concerns. This duality underscores the need for a careful consideration of privacy implications when designing and deploying audio classification systems. 

\section{Privacy-Aware Audio Classification: Challenges}
In this section, we discuss the major hurdles in designing privacy-preserving audio-based systems. 
\begin{itemize}
  \item \textbf{Expanding Privacy Beyond Speech}: Prior work mainly considers speech as the sensitive content and tries to filter out \cite{xia2020pams}, obfuscate \cite{liaqat2017method}, or replace \cite{chen2008audio} the speech segments. Audio contains more privacy invasive information than just speech. There is a need to look at a more holistic definition of privacy. 
  \item \textbf{Complexity of Privacy Evaluation}: Evaluating privacy in audio sensing systems is a non-trivial problem as privacy varies with context. Most common techniques are to use automatic speech recognition using Google APIs \cite{soundswordless} or perform human evaluation \cite{chhaglani2022flowsense}. Boovaraghavan  et al. \cite{boovaraghavan2024kirigami} use word error rate (WER) and phoneme error rate (PER) as the privacy evaluation metrics. However, these techniques do not evaluate whether the systems are leaking speaker related information or their location. For example, the system could infer myriads of details about an individual even with high WER. Thus, there is a need for accurate privacy evaluation mechanism.
  \item \textbf{Towards Universal Privacy Preservation}: Some of the prior work has designed privacy preserving pipelines for different applications. Bourlard et al. \cite{bourlardprivacy} uses privacy sensitive features for speech/non-speech detection, while Parthasarathi et al. \cite{parthasarathi2009speaker} uses privacy sensitive audio features to detect speaker change in conversations. Although, these techniques are useful, they are application specific and require manual effort to design hand-crafted features. There is a need to address the challenge of privacy from a more general perspective by designing a set of generalizable privacy-aware features.  
  \item \textbf{Privacy-Accuracy Trade-off}: The accuracy-privacy tradeoff in audio classification systems represents a critical challenge \cite{chhaglani2022flowsense}. On one hand, achieving high accuracy in audio classification often necessitates the use of detailed and comprehensive audio features, such as MFCCs. These features capture nuanced aspects of the audio signal, allowing deep learning models to make precise inferences about the content, context, or identity associated with the audio data. On the other hand, the very richness of information that enables such accuracy can also lead to significant privacy concerns, as these features may contain or enable the reconstruction of sensitive information about individuals or their environment.
\end{itemize}

To address these challenges, we propose to redefine privacy in the context of audio classification to encompass more than just speech. We aim to use \textit{privacy-preserving audio features that are universally applicable and maintain effectiveness of audio classification systems}.


\section{Preliminary Experiments with ESC-50}
\subsection{Dataset Description}
The ESC-50  \cite{piczak2015esc} dataset is a comprehensive and well-curated collection designed for the task of environmental sound classification, highlighting the diversity and complexity of auditory scenes that can be encountered in everyday life. ESC-50 is a labeled collection of 2,000 environmental audio recordings (each 5 seconds long) with 50 classes. These categories encompass a wide range of sounds from natural environments (such as rain, thunderstorms, and animal sounds), human-made noises (like car horns, alarms, and mechanical tools), and human sounds (including laughing, clapping, and crying). The dataset is fully annotated, with labels indicating the category of each audio clip. 
\subsection{Generalizable Privacy-Preserving Features}
We process the audio files using a sliding window of 500 ms with 100 ms overlap. We remove silent periods from the audio files using the top decibel threshold of 20. Next, each audio segment is converted into a list of time and frequency domain audio features. These features are as follows:
\begin{itemize}
    \item Zero Crossing Rate (ZCR) indicates the number of times the signal changes sign. This can be related to the texture of the sound, which is very relevant in differentiating between diverse environmental sounds.
    \item Harmonic-to-Noise Ratio (HNR) measures the amount of harmonic content compared to noise within a signal. Since environmental sounds can be either more harmonic (e.g., bird singing) or more noisy (e.g., rain), this feature can help distinguish between them.
    \item Spectral Contrast refers to the difference in amplitude between peaks and valleys in the sound spectrum.
    \item Peak, RMS, Energy relate to the amplitude and power of the audio signal and are useful for distinguishing between loud and soft sounds, as well as the intensity of the sound source.
    \item Spectral Roll-off indicates the frequency below which a certain percentage of the total spectral energy is contained. This feature helps to separate sounds with high-frequency content from those with low-frequency content.
    \item Spectral Flatness and Bandwidth (BW) provide a measure of how noise-like a sound is, versus being tonal. 
    \item Spectral Centroid is a measure of the 'brightness' of a sound, which can be useful to characterize sounds with high-frequency content like bird songs or bells.
\end{itemize}
We do not include features that contain privacy invasive information. For instance, fundamental frequency can reveal details about speaker or formants can reveal speech content as mentioned in source filter model of speech production \cite{fant1981source}.

\subsection{Classification Results}
\begin{figure}
    \centering
    \includegraphics[width=0.78\columnwidth]{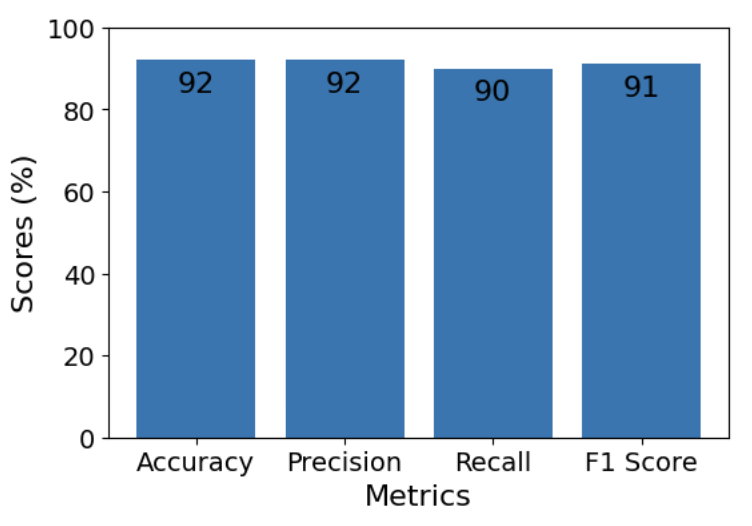}
    \caption{Classification metrics using privacy-preserving features}
    \label{fig:metrics}
\end{figure}
We use these features to train a random forest classifier model. To evaluate the model, we split the dataset into train, test and validation set (70\%, 15\%, 15\%) after shuffling. We achieve aggregate accuracy of 92.23\% over all the classes. We observe the feature importances using gini importance aggregated over all the classes. We observe that spectral contrast, ZCR and HNR are the most important features for this task. This is because ZCR is particularly useful in distinguishing between percussive sounds (like footsteps or clapping) and more harmonic sounds (like animal calls or musical instruments). HNR is crucial for differentiating between sounds that are more tonal and structured, such as human voices or music, versus those that are more stochastic and noise-like, such as wind or rushing water.

\subsection{Comparison with Non Privacy-based approaches}
Most of the prior work uses MFCCs or the Mel spectrogram for classifying ESC-50 classes. These spectrograms can be use to infer human speech \cite{ittichaichareon2012speech} and speaker identity \cite{yutai2009speaker}. The accuracy results obtained from the ESC-50 dataset using spectrogram as feature input has a wide range of variation depending on the spectrogram type and the DL model. Mu et al. \cite{mu2021environmental} use Log-Mel spectrogram and temporal-frequency attention based convolutional neural network model (TFCNN), and achieve accuracy of 84.4\%. Wang et al. \cite{wang2019environmental} propose parallel temporal-spectral attention mechanism for CNN to learn discriminative time-frequency representations of the spectrogram. This model achieve accuracy of 88.6\%. Certain models \cite{gazneli2022end} achieve very high accuracy of >90\% using pre-training and data augmentation techniques.
Thus, we can achieve comparable or better accuracy by taking the privacy-aware route. 
The proposed features are generalizable as they can be used for 50 different environment sounds

\section{Conclusions}
This paper explores the privacy concerns in audio classification systems that use features like MFCCs and spectrograms, which can inadvertently reveal sensitive personal information beyond just speech content, such as speaker identity, location, and other metadata. We identify key open challenges including developing robust privacy evaluation mechanisms and navigating the accuracy vs. privacy tradeoff. To address these issues, we propose a set of generalizable privacy-preserving audio features and demonstrate their ability to achieve comparable accuracy to prior work on the ESC-50 environmental sound dataset, while mitigating privacy risks. Overall, this work represents an important step towards enabling privacy-preserving audio sensing systems that can unlock the potential of this rich data modality while effectively safeguarding user privacy.



\section*{Acknowledgments} 
This research was supported in part by NSF grants 2211302, 2211888, 2213636, 2105494, and  US Army contract W911NF-17-2-0196. Any opinions, findings, conclusions, or recommendations expressed in this paper are those of the authors and do not necessarily reflect the views of the funding agencies.

\bibliographystyle{unsrt}
\bibliography{citation}

\end{document}